\RequirePackage{fix-cm}
\documentclass[twocolumn,final]{svjour3}          
\smartqed  
\usepackage{graphicx}
\usepackage{amsmath}
\usepackage{setspace}
\usepackage{amsfonts}
\usepackage{mathtools}
\usepackage{subfig}
\usepackage{lineno}
\usepackage{algorithm}
\usepackage{algpseudocode}
\usepackage{hyperref}
\usepackage{float}
\usepackage{array,multirow}
\usepackage{color}
\usepackage{tikz}
\usepackage{graphicx}
\usepackage[flushleft]{threeparttable}
\usepackage{blindtext}
\usepackage[colorinlistoftodos,prependcaption,textsize=tiny]{todonotes}
\usepackage{regexpatch}
\usepackage[export]{adjustbox}
\usepackage{booktabs}
\usepackage[normalem]{ulem}

\usepackage{scalerel,amssymb}

\def\msquare{\mathord{\scalerel*{\Box}{gX}}}

\makeatletter
\xpatchcmd{\@todo}{\setkeys{todonotes}{#1}}{\setkeys{todonotes}{inline,#1}}{}{}
\makeatother

\newcommand{\mat}[1]{\mathsf{#1}} 
\newcommand{\vect}[1]{#1}

\newcommand{\eigfun}{\phi}

\newcommand{\eigval}{\lambda}
\newcommand{\eigvec}{\mat{v}^h}

\newcommand{\x}{\vect{x}}
\newcommand{\xp}{\vect{x}^\prime}
\newcommand{\px}{\hat{\x}}
\newcommand{\pxp}{\hat{\x}^\prime}

\newcommand{\dint}[1]{\,\mathrm{d}{#1}}

\newcommand{\midx}[1]{#1}

\newcommand{\bspline}[1]{B_{\midx{#1}}}
\newcommand{\nurbs}[1]{R_{\midx{#1}}}
\newcommand{\geomap}[1]{F(#1)}
\newcommand{\geomapNOARG}[1]{F}

\newcommand{\detgeomap}[1]{\mathrm{det}\,\mathrm D\geomap{#1}}

\newcommand{\tbspline}[2]{\tilde{B}_{\midx{#1}}(#2)}

\newcommand{\nomatA}{A}
\newcommand{\nomatB}{Z}
\newcommand{\matA}{\mat \nomatA}
\newcommand{\matB}{\mat \nomatB}

\newcommand{\stochdom}{\Theta}
\newcommand{\w}{\theta}
\newcommand{\sigfield}{\Sigma}
\newcommand{\pmeasure}{\mathbb P}

\newcommand{\covfun}[2]{\Gamma(#1,{#2})}

\newcommand{\randfield}{\alpha}

\newcommand{\randvar}{\xi}

\newcommand{\dom}{\mathcal D}
\newcommand{\pdom}{{\hat{\mathcal{D}}}}

\begin{document}

\title{A comparison of matrix-free isogeometric Galerkin and collocation methods for Karhunen--Loève expansion}


\author{Michal L. Mika \and Ren\'e R. Hiemstra \and Thomas J.R. Hughes \and Dominik Schillinger}


\institute{Michal L. Mika \at
              \email{mika@ibnm.uni-hannover.de}           
           \and
           René R. Hiemstra \at
              \email{rene.hiemstra@ibnm.uni-hannover.de}           
	   \and
	   Dominik Schillinger \at
              \email{schillinger@ibnm.uni-hannover.de}  \\
              Institute of Mechanics and Computational Mechanics\\
              Leibniz University Hannover\\
              Appelstr. 9a, 30167 Hannover, Germany        
               \and
           Thomas J.R. Hughes \at
           \email{hughes@oden.utexas.edu}\\
              Oden Institute for Computational Engineering and Science\\
              The University of Texas at Austin\\
              201 East 24th Street, C0200, Austin, TX 78712-1229 USA \\
}

\date{Received: date / Accepted: date}

\maketitle

\begin{abstract}
Numerical computation of the Karhunen--Lo\`eve expansion is computationally challenging in terms of both memory requirements and computing time. We compare two state-of-the-art methods that claim to efficiently solve for the K--L expansion: (1) the matrix-free isogeometric Galerkin method using interpolation based quadrature proposed by the authors in \cite{mika_matrix-free_2020} and (2) our new matrix-free implementation of the isogeometric collocation method proposed in \cite{jahanbin_isogeometric_2019}. Two three-dimensional benchmark problems indicate that the Gal\-erkin method performs significantly better for smooth covariance kernels, while the collocation method performs slightly better for rough covariance kernels.
\end{abstract}

\keywords{Karhunen--Loève expansion \and Galerkin \and collocation \and matrix-free \and isogeometric analysis}


\section{Introduction} \label{sec:introduction}
The Karhunen--Lo\`eve (K--L) expansion decomposes a random field into an infinite linear combination of $L^2$ orthogonal functions with decreasing energy content. Truncated representations have applications in stochastic finite element analysis (SFEM) \cite{keese_review_2003,stefanou_stochastic_2009,sudret_stochastic_2000}, proper orthogonal decomposition (POD) \cite{lu_review_2019,rathinam_new_2003} and in image processing where the technique is known as principal component analysis (PCA) \cite{jolliffe_principal_2016}. All these techniques are closely related and widely used in practice \cite{liang_proper_2002}. 

Numerical approximation of the K--L expansion by means of the Galerkin or collocation method leads to a generalized eigenvalue problem:
\textit{Find} $(\eigvec_k, \lambda^h_k) \in \mathbb{R}^N \times \mathbb{R}^+$ \textit{such that}
\begin{align}  \label{eq:matrix_equations} 
	& \matA \eigvec = \lambda^h_k \matB \eigvec &
	&\text{for } & &k=1,2, \ldots , M.&
\end{align} 
This matrix problem is computationally challenging for the following reasons: (1) the matrix $\matA$ is dense and thus memory intensive to store explicitly; (2) every iteration of an iterative eigenvalue solver requires a backsolve of a factorization of $\matB$; and (3) the assembly of $\matA$ is computationally expensive\footnote{Formation and assembly costs for a standard Galerkin method scale $\mathcal{O}(N^2_e \cdot  (p+1)^{3d}))$, where $N_e$ is the number of finite elements, $p$ is the polynomial degree and $d$ is the spatial dimension.}.

In this paper, we investigate and compare two state-of-the-art methods that were recently proposed to efficiently solve for  the K--L expansion. The first method is the matrix-free isogeometric Galerkin method proposed by the authors in~\cite{mika_matrix-free_2020}, which uses an advanced quadrature technique to gain high performance that is scalable with polynomial order. The second method is our new matrix-free implementation of the isogeometric collocation method proposed in~\cite{jahanbin_isogeometric_2019}. As a collocation method it requires far fewer quadrature points than a standard Ga\-lerkin method  such that the  assembly of the collocation equations is simple and efficient.

This paper is structured as follows. In Section \ref{sec:background}, we briefly review the 
basic aspects of the K--L expansion in the context of random field represenations. In Section \ref{sec:methods}, we concisely present the two matrix-free solution methods and assess their algorithmic complexity. Three-dimensional numerical benchmark problems with comparisons in terms of accuracy and solution time are provided in Section \ref{sec:examples}. We  summarize  our conclusions in Section \ref{sec:conclusions} and discuss future work.

\section{Karhunen--Loève expansion of random fields} \label{sec:background}
Consider a complete probability space $(\stochdom, \sigfield, \pmeasure)$ where $\stochdom$ denotes a sample set of random events and $\pmeasure$ is a probability measure $\pmeasure\, : \,\sigfield\rightarrow [0,1]$. Let $\randfield(\cdot ,  \w) \; : \; \stochdom \mapsto L^2(\dom)$  denote a random field on a bounded domain \linebreak $\dom \in \mathbb{R}^d$ with mean $\mu(\x) \in L^2(\dom)$ and covariance function $\covfun{\x}{\xp} \in L^2(\dom \times \dom)$. The K--L expansion of the random field $\randfield(\cdot, \w) $ requires the
solution of an integral eigenvalue problem. Consider the self-adjoint positive semi-definite linear operator $T \; :\; L^2(\dom) \mapsto L^2(\dom)$,
\begin{align}
\label{eq:hilbert_schmidt}
\quad \left(T  \eigfun \right) (\x) := \int_{\dom} \covfun{\x}{\xp}\eigfun(\xp) \dint{\xp}.
\end{align}
 The  eigenfunctions $\{\eigfun_i \}_{i\in \mathbb N}$ of $T$ are defined by the homogeneous Fredholm integral eigenvalue problem of the second kind,
\begin{equation}
	\label{eq:fredholm}
	T\eigfun_i = \eigval_i \eigfun_i, \quad \eigfun_i \in L^2(\dom) \text{ for } i \in \mathbb{N}.
\end{equation}
The eigenfunctions $\eigfun_i$ are orthonormal in $L^2(\dom)$ and the corresponding eigenvalues form a non-increasing  sequence $ \eigval_1 \geq  \eigval_2 \geq \cdots \geq  0$. The K--L expansion of the random field $\randfield(\cdot, \w) $ is given as
\begin{align}
	\randfield(\x, \w) &= \mu(\x) + \sum\limits_{i=1}^{\infty} \sqrt{\eigval_i} \eigfun_i(\x) \randvar_i(\w)
\label{eq:kle}
\end{align}
where
\begin{align}
	\randvar_i(\w) &:= \frac{1}{\sqrt{\eigval_i}} \int_{\dom} \left(\randfield(\x, \w) - \mu(\x) \right) \eigfun_{i}(\x)\dint{\x}.
\end{align}
Truncating the series in \eqref{eq:kle} after $M$ terms leads to an approximation of $\randfield$ denoted by $\randfield_M$. For practical computations in the context of stochastic finite element methods \cite{keese_review_2003,stefanou_stochastic_2009,sudret_stochastic_2000}, the truncation order $M$ is typically chosen between 20 and 30 terms \cite{eiermann_computational_2007,stefanou_stochastic_2009}. Each term in the expansion introduces one stochastic dimension, which is an example for the \textit{curse of dimensionality}.

\section{Numerical methods} \label{sec:methods} 
In this section we briefly review the matrix-free Ga\-lerkin method proposed in \cite{mika_matrix-free_2020} and  introduce our matrix-free implementation  of the  isogeometric  collocation me\-thod proposed in \cite{jahanbin_isogeometric_2019}. We include an analysis of the algorithmic complexity in terms of the polynomial degree $p$ and number of elements $N_e$ of the $d$-dimensional spatial domain $\dom$. 

In both approaches the generalized  algebraic  eigenvalue problem is  first reformulated as a standard algebraic eigenvalue problem using standard linear algebra techniques~\cite{saad_numerical_2011}: \textit{Find} $(\eigvec_k, \lambda^h_k) \in \mathbb{R}^N \times \mathbb{R}^+$ \textit{s.t.}
\begin{align} 
	\qquad \begin{cases}
		\matA^\prime \mat{v}_k^\prime = \lambda_k^h \mat v_k^\prime \\
		\eigvec_k = \mat{C} \mat{v}_k^\prime
	\end{cases}
	 \quad \text{for}\quad k = 1,2,\ldots, M .
\label{eq:standard}
\end{align} 
Here $\mat{C}$ is an invertible mapping that depends on $\matB$ and $\matA^\prime$ can be written in terms of $\matA$ and $\matB$. 

\subsection{Matrix-free isogeometric Galerkin method} \label{sec:ibq} 
A variational treatment of \eqref{eq:fredholm} leads to the following problem: \textit{Find} $(\eigfun, \lambda) \in L^2(\dom) \times \mathbb{R}^+$ \textit{s.t.} $\forall \psi\in L^2(\dom)$
\begin{align}
\int_{\dom} \left( \int_{\dom} \covfun{\x}{\xp}\eigfun(\xp) \dint{\xp}  - \lambda \eigfun(x) \right) \psi(x) \dint{\x} = 0 .
\label{eq:galerkin} 
\end{align} 

From equation \eqref{eq:galerkin}, the Galerkin method is obtained by replacing $\eigfun, \psi \in L^2(\dom)$ by finite dimensional representations $\eigfun^h, \psi^h \in \mathcal{S}^h \subset L^2(\dom)$.  Being  posed in the variational setting, Galerkin methods inherit several advantageous properties such as exact $L^2$ orthogonality of the numerical eigenvectors and monotonic convergence of the numerical eigenvalues \cite{atkinson_numerical_1997,ghanem_stochastic_1991}. Furthermore, powerful tools exist in the variational setting to study the stability and convergence of the method\footnote{In general the stability and convergence analysis are challenging in the context of collocation methods.}.

With a trial space $\mathcal S^h := \mathrm{span} \left\{ N_{i}(\x) \right\}_{i = 1,\ldots,N}$ the Galerkin method leads to the eigenvalue problem defined in \eqref{eq:matrix_equations} with the system matrices%
\begin{subequations}
	\label{eq:galerkin_system_matrices}
\begin{align}
	\nomatA_{ i j} &:= \int_{\dom} N_{i}(\x)  \int_{\dom} \covfun{\x}{\xp} N_{j}(\xp) \dint{\xp} \dint{\x} 	 			\label{eq:galerkin_system_matrices_1}				\\
	\nomatB_{ i j} &:= \int_{\dom} N_{i}(x) N_{j}(x) \dint{\x}
	\label{eq:galerkin_system_matrices_2}
\end{align} 
\end{subequations}
Alternatively, the eigenvalue problem can be solved in the standard form introduced in equation \eqref{eq:standard} where $\matA^\prime := \mat{L}^{-1} \matA \mat{L}^{-\top}$ and $\mat{C} := \mat{L}^{-\top}$. The matrix $\mat{L}$ is defined by the lower triangular matrix in the Cholesky decomposition of $\matB = \mat{L} \mat{L}^{\top}$.

Typically, the space  $\mathcal S^h$ is spanned by piecewise $C^0$-continuous polynomial functions on quadrilateral, he\-xagonal or simplicial elements \cite{ghanem_stochastic_1991}. Recently, non-uni\-form rational B-splines (NURBS) have been applied in the context of an isogeometric Galerkin method~\cite{rahman_galerkin_2018}. These methods commonly evaluate the integrals in \eqref{eq:galerkin_system_matrices} using standard numerical quadrature rules. A Gauss--Legendre numerical qua\-drature rule leads, however, to an algorithmic complexity of $\mathcal{O}(N_e^2 \cdot  (p+1)^{3d})$ \cite{mika_matrix-free_2020}, which becomes  excessively expensive with the number of elements $N_e$, polynomial degree $p$ and spatial dimension~$d$. Furthermore, as mentioned in the introduction, the matrix $\matA$ is dense and requires  $\mathcal O(8 \cdot N^2)$  bytes of storage in double precision  arithmetic, where $N$ is the number of degrees of freedom in the trial space.

To overcome these limitations, the matrix-free Ga\-lerkin method proposed in \cite{mika_matrix-free_2020} avoids storing  the main system matrix $\matA$ and achieves computational efficiency by utilizing a non-standard trial space in combination with a specialized quadrature technique, called \emph{interpolation based quadrature}. This approach requires a minimum number of quadrature points and enables application of global sum factorization techniques \cite{bressan_sum_2019}. We sketch the main ideas of the method and refer to \cite{mika_matrix-free_2020} for further details. 

Let  $\{ \bspline{i}(\px) \}_{i = 1,\ldots,N}$ and $\{ \tbspline{j}{\px} \}_{j = 1,\ldots,\tilde{N}}$ denote two sets of tensor product B-splines of, for simplicity, uniform polynomial degree $p$. The first set is used in the definition of the trial space, whereas the second set is used in a projection of the kernel $\Gamma(\x,\xp)$  and is a  part of the \emph{interpolation based quadrature}. Let $F : \pdom \rightarrow \dom$ be the geometric mapping from the reference domain to the physical domain. The trial space is defined as
\begin{equation} 
	\mathcal S^h := \mathrm{span} \left\{ \bspline{i}(\px) / \sqrt{ \detgeomap{\px}} \right\}_{i = 1,\ldots,N.} \label{eq:trialspace} 
\end{equation} 
The advantage of this particular choice of  the  trial space is that the mass matrix in \eqref{eq:galerkin_system_matrices_2} has  a  \emph{Kronecker} structure and can be factored as $\matB = \mat{Z}_d \otimes \cdots \otimes \mat{Z}_2 \otimes \mat{Z}_1$, where $ \{ \mat{Z}_k \}_{k=1,2,\ldots,d}$ are univariate mass matrices.  By leveraging this factorization  the matrix-vector products of Kronecker matrices can be evaluated in nearly linear time complexity. This also holds for the matrix $\mat{L}$ in the Cholesky factorization of $\matB$, which is factored as $\mat{L} = \mat{L}_d \otimes \cdots \otimes \mat{L}_2 \otimes \mat{L}_1$ from which the respective inverse follows as $\mat{L}^{-1} = \mat{L}_d^{-1} \otimes \cdots \otimes \mat{L}_2^{-1} \otimes \mat{L}_1^{-1}$.

The  interpolation based quadrature in combination with  the  choice of the trial space in \eqref{eq:trialspace} leads to a factorization of  the  matrix $\matA$ as
\begin{align}
	\matA = \mat{M}^\top \mat{\tilde B}^{-1} \mat{J} \mat{\Gamma} \mat{J} \mat{\tilde B}^{-\top} \mat{M}.
\end{align} 
Here $\mat{\Gamma} := \Gamma(\x_i , \x_j) \in \mathbb{R}^{\tilde{N} \times \tilde{N}}$ is the covariance kernel evaluated at the Greville abscissae, $\mat{J} \in \mathbb{R}^{\tilde{N} \times \tilde{N}}$ is the square root of a diagonal matrix  of determinants of the Jacobian of the mapping at these points  and the matrices $\mat{\tilde B} = \mat{\tilde B}_d \otimes \cdots \otimes \mat{\tilde B}_2 \otimes \mat{\tilde B}_1 \in \mathbb{R}^{\tilde{N} \times \tilde{N}}$ and $\mat{M} = \mat{M}_d \otimes \cdots \mat{M}_2 \otimes \mat{M}_1 \in \mathbb{R}^{\tilde{N} \times N}$ are Kronecker product matrices. In fact $\mat{\tilde B}_k$ and $\mat{M}_k$, $k=1,2,\ldots,d$, are univariate collocation and mass matrices, respectively, which are introduced by the interpolation based quadrature. The computation of the eigenvalues and eigenvectors requires evaluation of matrix-vector products $\mat v^\prime \mapsto \matA^\prime \mat v^\prime$. This leads to a nine step algorithm presented in \cite{mika_matrix-free_2020}. The matrix-vector products with the Kronecker structured matrices $\mat{L}^{-\top}$, $ \mat{M}$, $\mat{B}^{-\top}$ and the diagonal matrix~$\mat{J}$ as well as all the respective transpose operations are performed in linear or nearly linear time complexity. The matrix-vector products with the matrix $\mat{\Gamma}$ are performed in quadratic time complexity. Hence, our matrix-free algorithm scales  quadratically  with the dimension of the interpolation space $\tilde{N}$. We note that in this algorithm, the matrix rows of~$\mat{\Gamma}$ are computed on the fly, which saves memory by not explicitly storing the dense matrix $\mat{\Gamma}$. Memory requirements for the remaining matrices are negligible, since they are either diagonal or Kronecker product matrices. For additional details about the matrix-free method, interpolation based quadrature and Kronecker products, we refer to \cite{mika_matrix-free_2020}.

\subsection{Matrix-free isogeometric collocation method} \label{sec:colloc} 
In contrast to a Galerkin method, a collocation method does not treat the integral equation \eqref{eq:fredholm} in a variational manner. Instead, we require the discretized residual 
\begin{equation}
r^h(\x) := \int_{\dom} \covfun{\x}{\xp}\eigfun^h(\xp) \dint{\xp} - \lambda^h \eigfun^h(\x) 
\label{eq:colloc_residual} 
\end{equation} 
to vanish at distinct points $\x \in \dom$. In \cite{jahanbin_isogeometric_2019}, the geometry and trial spaces are  discretized in terms of NURBS basis functions
\begin{equation}
	\mathcal{S}^h := \{\nurbs{i}(x)\}_{i = 1,\ldots,N}
\end{equation} 
in the sense of the isoparametric approach of isogeometric analysis. 
In this study, we choose to collocate~\eqref{eq:colloc_residual} at the Greville abscissae $\{\x_i \}_{i = 1,\ldots,N}$. The method is expressed concisely in matrix form \eqref{eq:matrix_equations} where the corresponding system matrices are given by 
\begin{align} \label{eq:colloc_system_matrices} 
	\nomatA_{ i j} := \int_{\dom} \covfun{\x_i}{\xp} \nurbs{j}(\xp) \dint{\xp}, \qquad \nomatB_{i j} := \nurbs{j}(\x_i). 
\end{align} 
In primal form \eqref{eq:standard}, this means that $\matA^\prime = \matB^{-1} \matA$ and $\mat{C}$  is the identity matrix.  The matrices $\mat A$ and $\mat Z$ are square and, in general, not symmetric. In contrast to variational methods, where the system matrices are symmetric and positive (semi)-definite by construction, collocation methods do not ensure a real-valued eigensolution for any element size $h>0$. For an in-depth exposition of the collocation method, we refer the reader to~\cite{atkinson_numerical_1997}, and to \cite{auricchio_isogeometric_2010,Schillinger:13.2} for details on the isogeometric formulation.

The matrix-free version of the collocation method is derived analogously to the matrix-free Galerkin method described above. Due to the properties of the system matrix $\matB$, instead of the Cholesky decomposition employed in the Galerkin me\-thod, we use the \textit{pivoted} $\mat L \mat U$ decomposition, $\mat P \matB \mat Q = \mat L\mat U$, to arrive at the standard matrix form. We observed that without pivoting the matrix-free collocation method suffers from numerical instabilities at polynomial orders $p>3$. We use the pivoted $\mat L\mat U$ decomposition of $\matB$ to apply the inverse of~$\matB$ to the matrix $\mat A$ and thus obtain~$\mat A^\prime$. The standard algebraic eigenvalue problem is then given by 
\begin{equation}
	\matA^\prime \mat v^\prime = \lambda \mat v^\prime \quad\text{where}\quad \matA^\prime := \mat Q \mat U^{-1} \mat L^{-1} \mat P \matA  
\end{equation} 
Following \cite{mika_matrix-free_2020}, we choose a row-wise evaluation of the coefficient vector in the standard matrix-vector product $\mat v^\prime \mapsto \matA^\prime \mat v^\prime$. The optimal evaluation order and further details for each step are given in Algorithm \ref{alg:mf_eval_colloc}.%
\begin{algorithm}[H]
\textbf{Input}:
	$v_{\midx j} \in \mathbb{R}^{N}$,
	$R_{\midx j \midx k} \in \mathbb{R}^{N \times (N_e\cdot N_q)}$,
	$P_{\midx i\midx j} $,
	$Q_{\midx i\midx j} $,
	$U_{\midx i\midx j} $,
	$L_{\midx i\midx j} \in \mathbb{R}^{N\times N}$,
	$J_{\midx k} \in \mathbb{R}^{N_e\cdot N_q}$,
	$W_{\midx k} \in \mathbb{R}^{N_e\cdot N_q}$\\
\textbf{Output}: $v^\prime_{\midx i} \in \mathbb{R}^{N}$
\begin{algorithmic}[1]
	\State $y_{\midx k} 		\leftarrow R_{\midx j \midx k} v_{\midx j}$		\Comment{Interpolation at quadrature points } 
	\State $y^\prime_{\midx k} 	\leftarrow y_{\midx k} \odot J_{\midx k} \odot W_{\midx k}$ \Comment{Scaling at quadrature points}
	\State $z_{\midx l} 		\leftarrow \Gamma_{\midx l\midx k} y^\prime_{\midx k}$	\Comment{Kernel evaluation one row at a time}
	\State $v^\prime_{\midx i} 	\leftarrow Q_{\midx i\midx t} U^{-1}_{ \midx t \midx r} L^{-1}_{\midx r \midx s} P_{\midx s\midx l}  z_{\midx l}$	\Comment{Backsolve using $\mat{LU}$ of $\matB$}
\caption{Matrix-free evaluation of the matrix-vector product $\mat{v}^\prime \mapsto \matA^\prime \mat{v}^\prime $ emerging from collocation}\label{alg:mf_eval_colloc}
\end{algorithmic}
\end{algorithm}
\subsection{Algorithmic complexity} \label{sec:complexity} 

\paragraph{Matrix-free Galerkin method}
Under the assumption of $\tilde{N} \propto N$, the formation and assembly costs are negligible compared to the matrix-vector products that scale independently of $p$ as $\mathcal O(\tilde{N}^2)$ \cite{mika_matrix-free_2020}. The total cost of the method scales as $\mathcal O(N_{\text{iter}} \cdot \tilde{N}^2)$, where $N_{\text{iter}}$ is the number of iterations of the eigenvalue solver.

\paragraph{Matrix-free collocation method}
We are interested in the algorithmic complexity of an element-wise assembly procedure for the system matrices that arise from the collocation method. We assume that (1) $\pdom$ has $N_e$ elements; (2) the products on every $d$-dimensional element~$\msquare^d$ in $\pdom$ are integrated with a quadrature rule $Q(f) := \sum_{k=1}^{N_q} w_k f(x_k)$ with $1 \leq N_q  \leq (p+1)^d$ quadrature points; and (3) the number of collocation points $N_c$ is equal to the number of degrees of freedom~$N$. The leading term in the total cost of formation and assembly arises from the cost of forming the element matrices,
\begin{align*} 
\nomatA^e_{\midx i\midx j} &= \int_{\msquare^d} \covfun{\px_\midx i}{\pxp} \nurbs{\midx j}(\pxp) \dint{\pxp}\\ & \approx  \sum_{k=1}^{N_q}  w_k \covfun{\px_\midx i}{\pxp_k} \bspline{\midx j}(\pxp_k) = \mat{C}_{\midx ik}  \mat{D}_{k\midx j} \\ &\quad \text{where}\quad {C}_{\midx i k} = w_k \covfun{\px_\midx i}{\pxp_k}\quad\text{and}\quad D_{k\midx j} = \nurbs{\midx j}(\pxp_k) 
\end{align*} 
with $\midx i = 1,\ldots,N$ and $j=1,\ldots, (p+1)^d$. The formation cost of $\mat C$ and $\mat D$ is negligible. The matrix-matrix pro\-duct cost is of $\mathcal O\left(N_c N_q (p+1)^d\right)$ and the cost for summation over all $N_e$ is of $\mathcal O\left(N_e N_c N_q (p+1)^d\right)$. Now, assuming a Gauss--Legendre quadrature rule with $N_q := (p+1)^d$ quadrature points and the proportionality relationship $N_e \propto N$, a collocation method with $N_c = N$ has a leading cost of $\mathcal O\left(N^2(p+1)^{2d}\right)$.

The algorithmic complexity in the matrix-free formulation is driven by the most expensive steps in Algorithm~\ref{alg:mf_eval_colloc}. In a single iteration of the eigenvalue solver, steps 1 and 3 have a complexity $\mathcal O(N\cdot N_e \cdot N_q)$. The element-wise multiplication in step 2 scales linearly with the number of quadrature points, $\mathcal O(N_e\cdot N_q)$. The last step scales as $\mathcal O(N^2)$. Evidently, steps 1 and 3 depend on the number of quadrature points. Since $N_e \cdot N_q \geq N$, they determine the overall cost of the method. Assuming a Gauss-Legendre quadrature rule with $N_q := (p+1)^d$ quadrature points in each element and $N_e \propto N$, the leading cost of a single iteration of the eigenvalue solver is $\mathcal O(N^2 (p+1)^d)$. Hence, the total cost of the matrix-free isogeometric collocation method scales as $\mathcal O(N_{\text{iter}}  \cdot N^2 (p+1)^d)$, where $N_{\text{iter}}$ is the number of iterations of the eigenvalue solver.

\paragraph{Comparison}
Compared to the matrix-free Galerkin me\-thod with interpolation based quadrature, the collocation method scales unfavourably with the polynomial degree. Furthermore, due to the lack of Kronecker structure, it is necessary to compute the pivoted $\mat L \mat U$ decomposition of the full matrix $\matB$. The computational cost of this factorization increases with $N$ as well as $p$, which is due to an increasing bandwidth of the matrix $\matB$.

\begin{remark} If the trial space in the collocation method is based on tensor product B-splines instead of NURBS, then the matrix $\matB$ is also a Kronecker product matrix, alleviating the disadvantage at large $N$ and $p$.
\end{remark}


\section{Numerical examples} \label{sec:examples} 
In this section, we compare  the  accuracy and efficiency of the matrix-free isogeometric Galerkin and collocation methods. In \cite{mika_matrix-free_2020}, it was shown that the proposed Galerkin method performed especially well in the case  of a smooth covariance kernel. For rough kernels, such as the $C^0$ exponential kernel, the interpolation based quadrature performed suboptimally. 




\begin{figure} 
\centering 
\includegraphics[width=0.7\columnwidth]{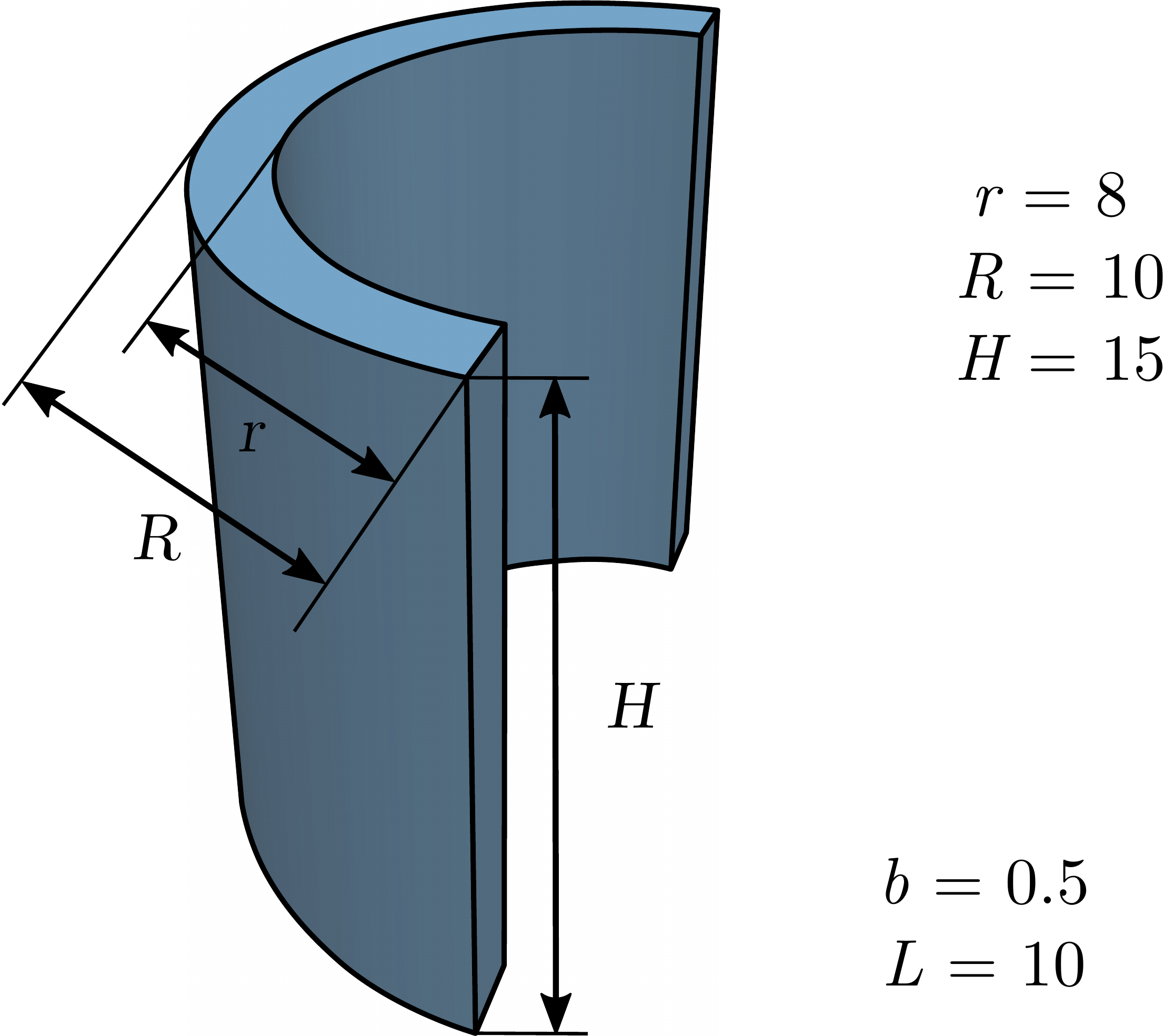} 
\caption{Benchmark geometry of a half-cylinder. The correlation length $bL=5$ is used throughout all cases.} 
\label{fig:model} 
\end{figure} 

In our study, we benchmark both methods for two kernels of different smoothness and appropriate refinement strategies of the spaces involved: (1) the exponential kernel together with $h$-refine\-ment and (2) the Gaussian kernel and $k$-refinement. In both variants, the solution space is equal for the Galerkin and collocation methods. The interpolation space used in the Galerkin method is defined on the same mesh as the solution space, but its continuity is adapted in accordance with the remarks made in \cite{mika_matrix-free_2020}.
All computations are performed sequentially on a laptop machine with an Intel(R) Core(TM) i7-9750H CPU @ 2.60GHz as well as 2x16 GB of DDR4 2666MHz RAM. Our reference solution is the standard isogeometric Galerkin solution computed on the finest possible mesh with a runtime of roughly 17 hours, tabulated in \cite{mika_matrix-free_2020}. 



 %

\subsection{Exponential covariance kernel}
In Example 1, we compare the performance with respect to $h$-refinement assuming an exponential kernel on the half-cylindrical domain shown in Figure \ref{fig:model}. 
The polynomial order in each parametric direction is $p=2$. We choose a tensor product Gauss--Legendre quadrature rule with $(p+1)^3$ points per element of the domain in the collocation method. In accordance with remarks made in \cite{mika_matrix-free_2020} the continuity of the interpolation space of the Galerkin method at the element interfaces is reduced to $C^0$. Furthermore, at element interfaces where the geometry is $C^0$, the interpolation space of the Galerkin method is set to $C^{-1}$.

\begin{figure}[htb!]
\centering 
\includegraphics[width=\columnwidth]{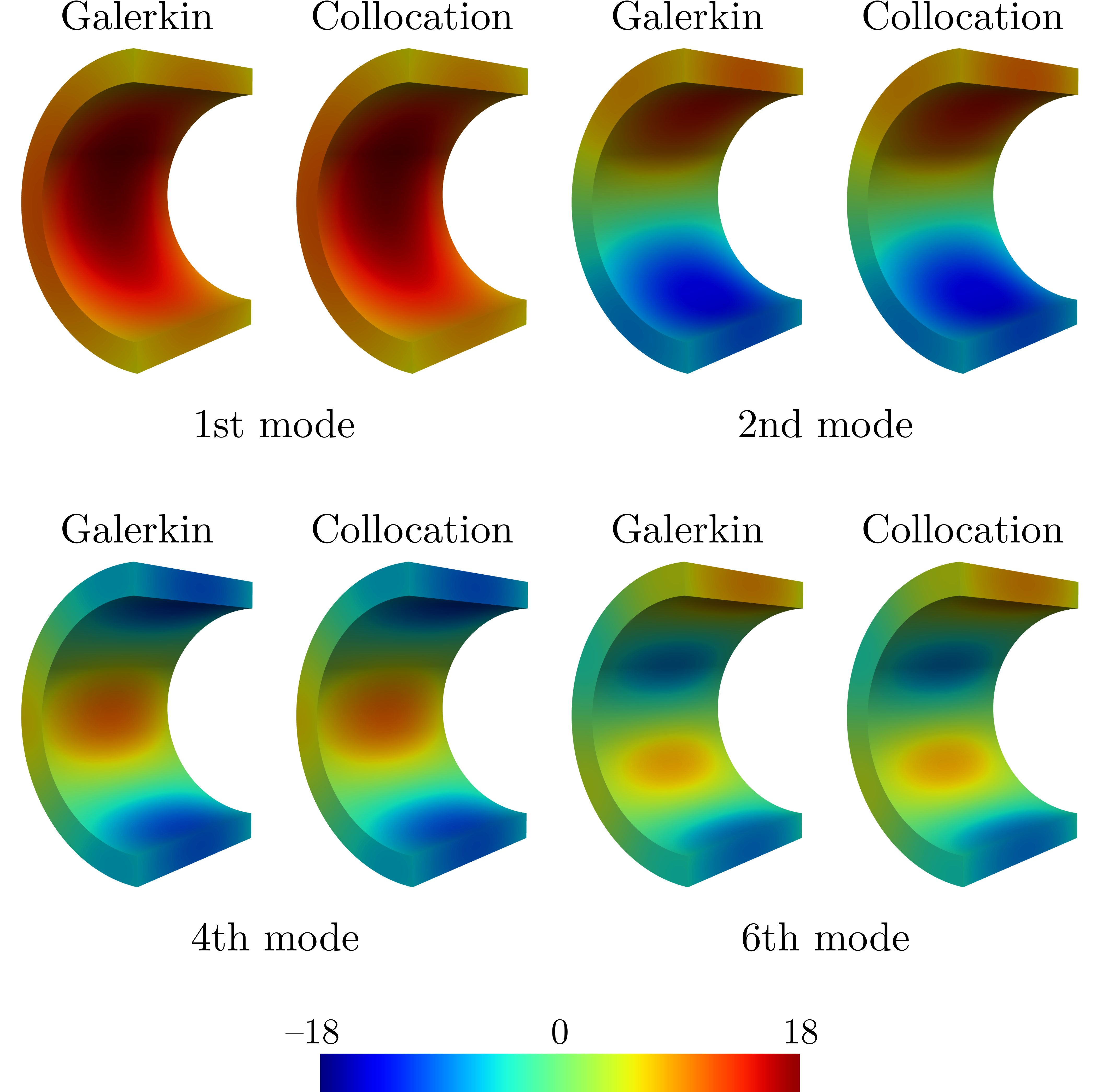} 
\caption{First, second, fourth and sixth eigenfunctions (Example 1, Case 1).\\} \label{fig:eigenmodes}
\includegraphics[width=0.95\columnwidth]{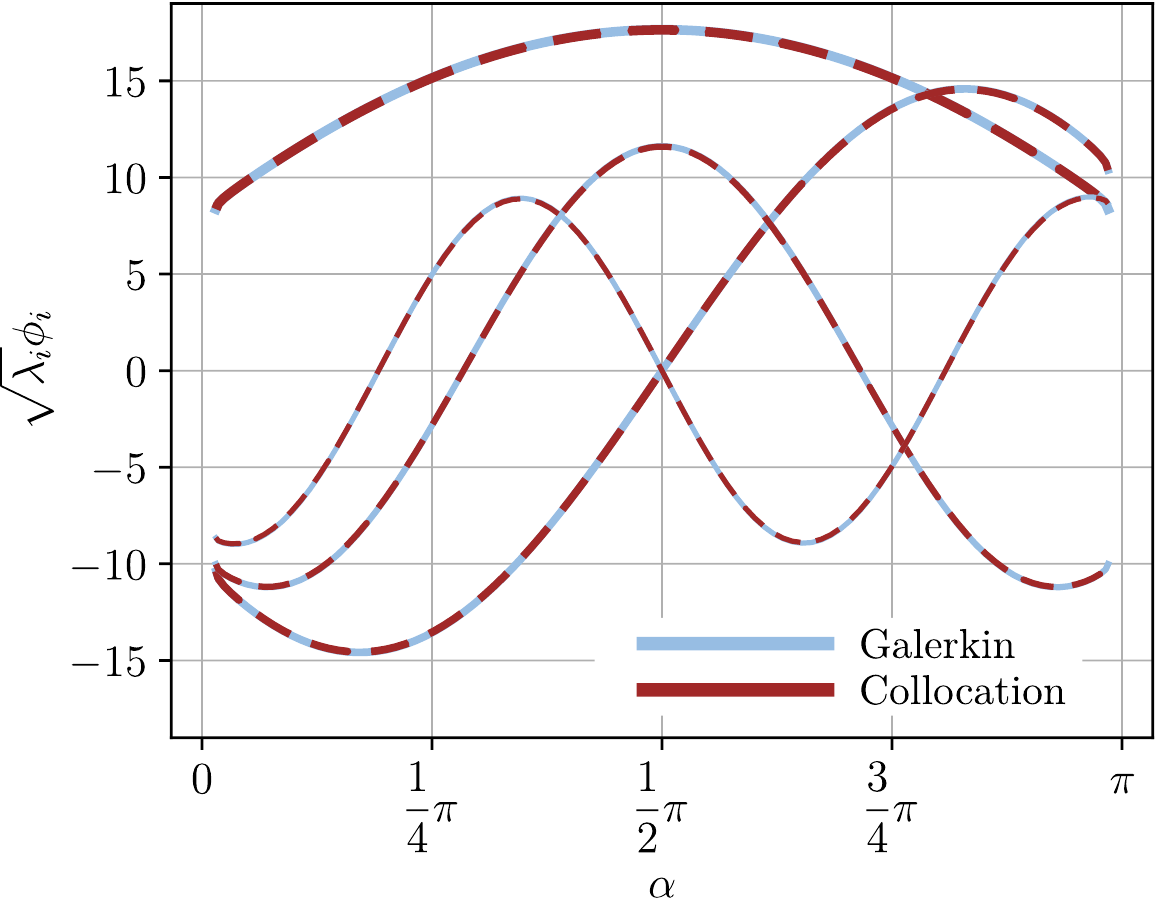} 
\caption{Line plot in the circumferential direction at the mid-planes of eigenfunctions in Figure~\ref{fig:eigenmodes}. Line-width decreases with increasing mode number.} 
\label{fig:eigenmodes-lineplot} 
\end{figure}
Our comparative investigation is based on five different resolution cases with respect to the characteristic size $h$ of the solution and interpolation mesh. Our specific choices of mesh size and number of degrees of freedom in the interpolation and solution spaces are summarized in Table \ref{tab:exp}. 
 \begin{table}[h!]
\caption{Mesh, solution space and interpolation space details in Example 1.} 
\label{tab:exp} 
\begin{tabular}{cccccc} 
\hline\noalign{\smallskip} & Case 1 & Case 2 & Case 3 & Case 4 & Case 5  \\ \noalign{\smallskip}\hline\noalign{\smallskip} $h$        & 2.857  & 1.719  & 1.556  & 1.423  & 1.142    \\ $N$        & 1050   & 2108   & 2800   & 3772   & 5625     \\ $\tilde N$ & 1980   & 8990   & 12210  & 16770  & 28294    \\ \noalign{\smallskip}\hline 
\end{tabular} 
\scriptsize{ \vspace{0.5em}\\ $h\,\,$ mesh size in the solution and interpolation mesh\\ $N\,\,$ number of degrees of freedom (dof) in the solution space\\ $\tilde N\,\,$ number of dof in the interpolation space (IBQ-Galerkin only)}
\end{table}

For Case 1, we visualize the first, second, fourth and sixth eigenfunctions computed by both methods, plotted in Figure \ref{fig:eigenmodes} on the half-cylinder domain. Figure~\ref{fig:eigenmodes-lineplot} illustrates that already for the coarsest resolution, both methods produce results that are practically indistinguishable from each other when plotted along a selected cut line.

For a quantitative comparison, let us introduce a relative eigenvalue error $\varepsilon_i$ with respect to the reference solution as
\begin{equation} 
	\varepsilon_i := \varepsilon(\lambda^{\mathrm{ref}}_{i}, \lambda_i^h) := \frac{\lvert \lambda^{\mathrm{ref}}_{i} - \lambda_i^h \rvert}{\lambda^{\mathrm{ref}}_{i}} 
\end{equation}
as well as a mean relative eigenvalue error $\overline{\varepsilon}$ given by 
\begin{equation} 
	\overline{\varepsilon} := \frac 1 M \sum\limits_{i=1}^{M} \varepsilon_i = \frac 1 M \sum\limits_{i=1}^{M} \frac{\lvert \lambda^{\mathrm{ref}}_{i} - \lambda_i^h \rvert}{\lambda^{\mathrm{ref}}_{i}}. 
\end{equation}

\begin{table}[h]
\caption{Color-coding to differentiate between five different cases and two different methods.} 
\centering
 \includegraphics[width=0.85\columnwidth]{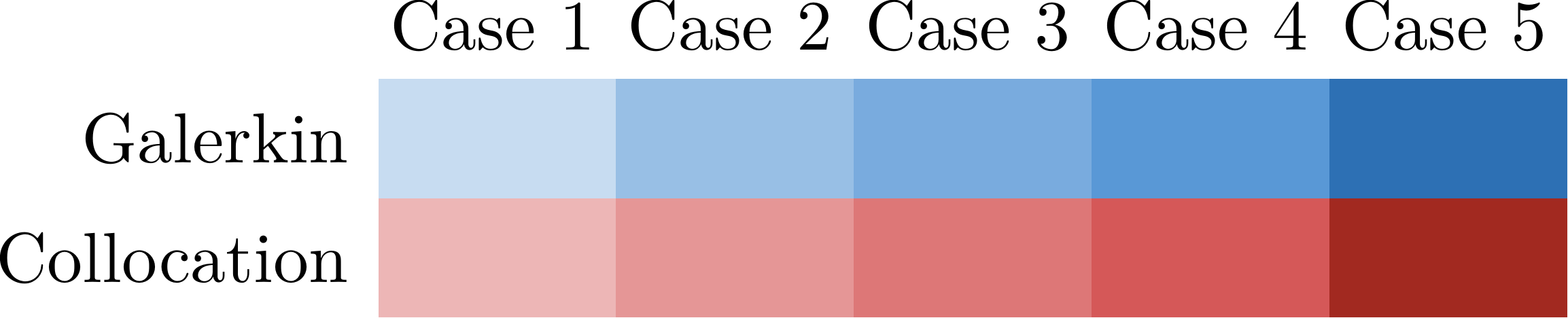} 
\label{tab:colorcoding} 
\end{table} 
To enable a concise illustration with respect to the five cases defined in Table \ref{tab:exp}, we define the color coding shown in Table~\ref{tab:colorcoding}. Blue indicates results obtained with the Galerkin method, red indicates results obtained with the collocation method. The change in shading from light to full color indicates the increasing mesh resolution from Case 1 to Case 5.


\begin{figure}[b!]
\centering 
\includegraphics[width=0.98\columnwidth]{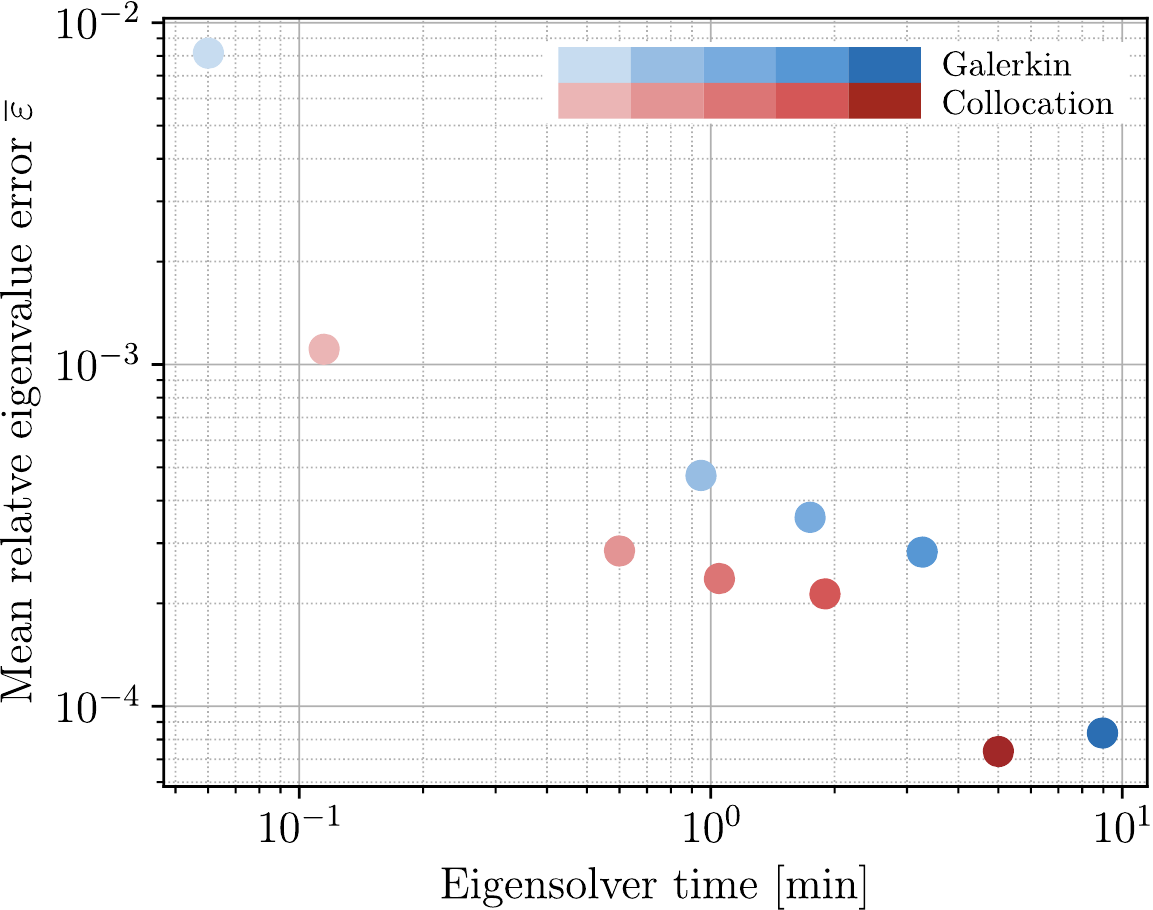} 
\caption{Mean relative eigenvalue error computed with the first 20 eigenvalues versus the eigensolver time (Example 1, exponential kernel).} 
\label{fig:scatter-exp} 
\vspace{0.16cm}
\includegraphics[width=\columnwidth]{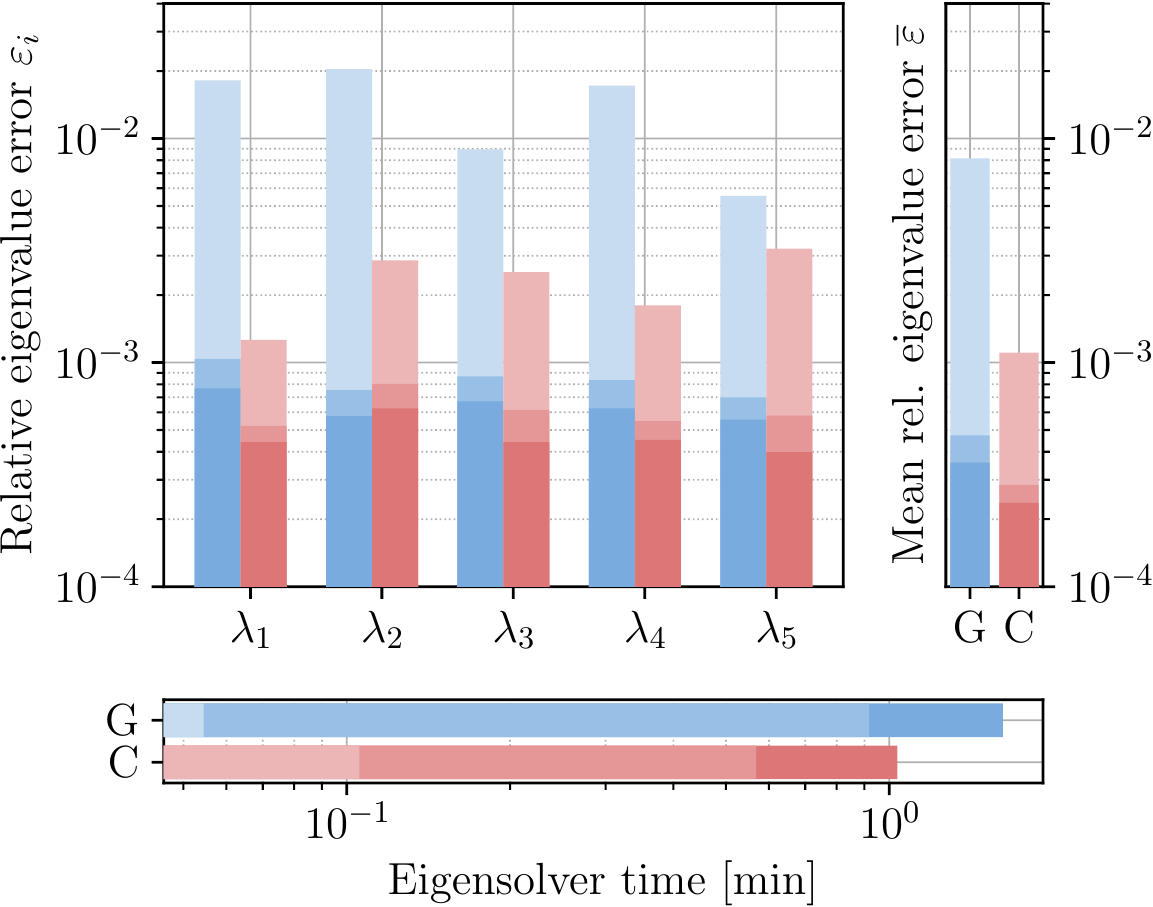} 
\caption{Error of the first five eigenvalues plotted for Cases 1--3 and corresponding timings and accuracy over the first 20 eigenvalues (Example 1, exponential kernel).} 
\label{fig:detail-exp}
\end{figure}%

Figure \ref{fig:scatter-exp} depicts relative accuracy versus computational time of the iterative eigensolver for the first twen\-ty eigenvalues measured against the reference solution. We observe that the collocation method performs roughly twice as fast at the same level of accuracy. 

In Figure~\ref{fig:detail-exp}, we present a detailed assessment of the accuracy of the first five eigenvalues. In addition, we provide an alternative visualization of the timings and the error in the first twenty eigenvalues.

\subsection{Gaussian covariance kernel}
In Example 2, we compare both methods for a smooth Gaussian covariance kernel. Since the integrand is\linebreak smooth, we expect that optimally smooth approximation spaces work best. Therefore, we fix the polynomial order $p$ and refine the approximation spaces with $C^{p-1}$ continuity between elements until a target mesh size of $2.857$ is reached ($k$-refinement). The resulting five different cases are summarized in Table \ref{tab:gau}. 
\begin{table}[h!]
\caption{Mesh, solution space and interpolation space details in Example 2.} 
\label{tab:gau} 
\begin{tabular}{cccccc} \hline\noalign{\smallskip} & Case 1 & Case 2 & Case 3 & Case 4 & Case 5   \\ \noalign{\smallskip}\hline\noalign{\smallskip} $p$        & 2      & 3      & 4      & 5      & 6        \\ $N$        & 1050   & 1628   & 2340   & 3198   & 4214    \\ $\tilde N$ & 1080   & 1672   & 2400   & 3276   & 4312     \\ \noalign{\smallskip}\hline 
\end{tabular} 
\scriptsize{ \vspace{0.5em}\\ $p\,\,$ polynomial order of the solution and interpolation space\\ $N\,\,$ number of degrees of freedom (dof) in the solution space\\ $\tilde N\,\,$ number of dof in the interpolation space (IBQ-Galerkin only)} 
\end{table}

Comparing Case 1 in Example 1 with Case 1 in Example 2, we find that the number of degrees of freedom in the interpolation space is smaller. This is due to the increased continuity at element interfaces of the interpolation space of the Galerkin method. This trend is also characteristic for \emph{k}-refinement and is observable in the remaining Cases 2--5.

\begin{figure} 
\centering 
\includegraphics[width=\columnwidth]{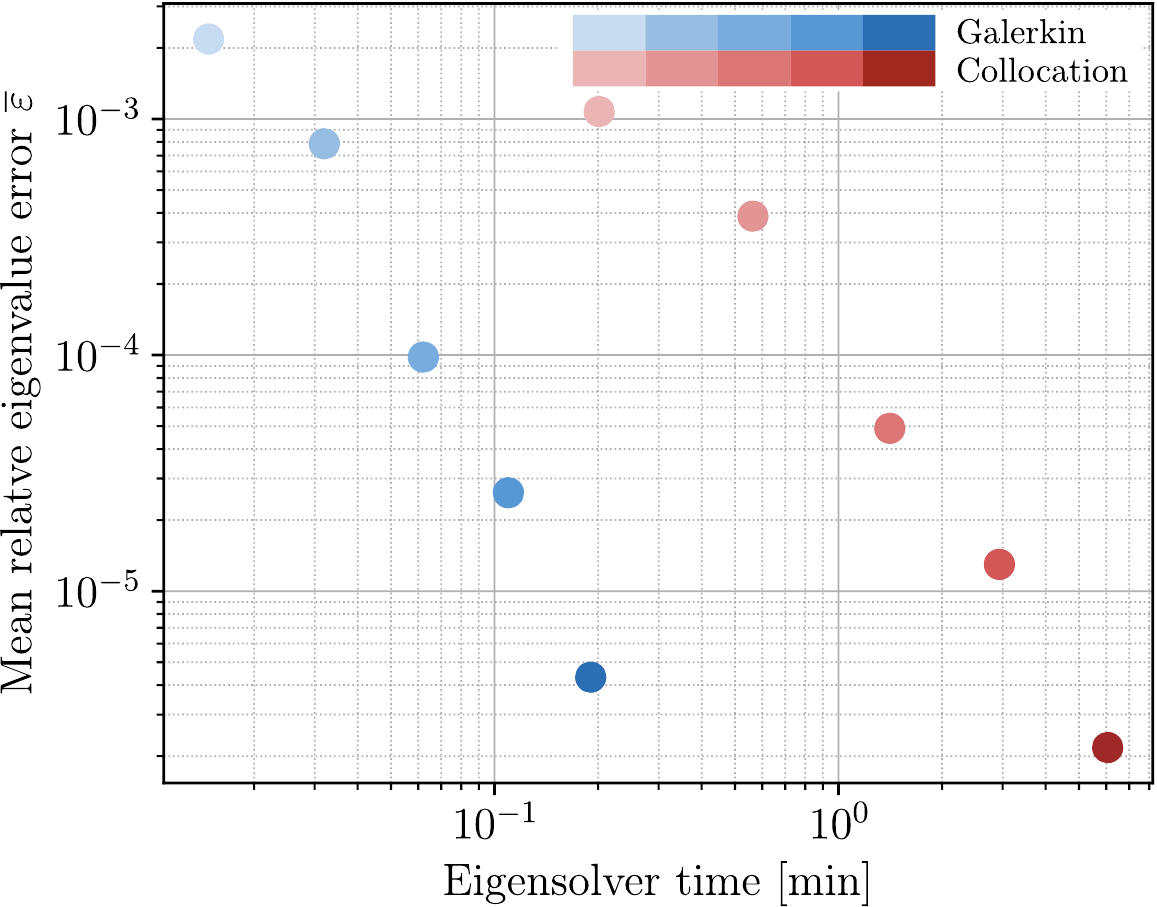} 
\caption{Mean relative eigenvalue error computed with the first 20 eigenvalues versus the eigensolver time (Example 2, smooth Gaussian kernel).} 
\label{fig:scatter-gau} 
\end{figure}

\begin{figure} 
\centering 
\includegraphics[width=\columnwidth]{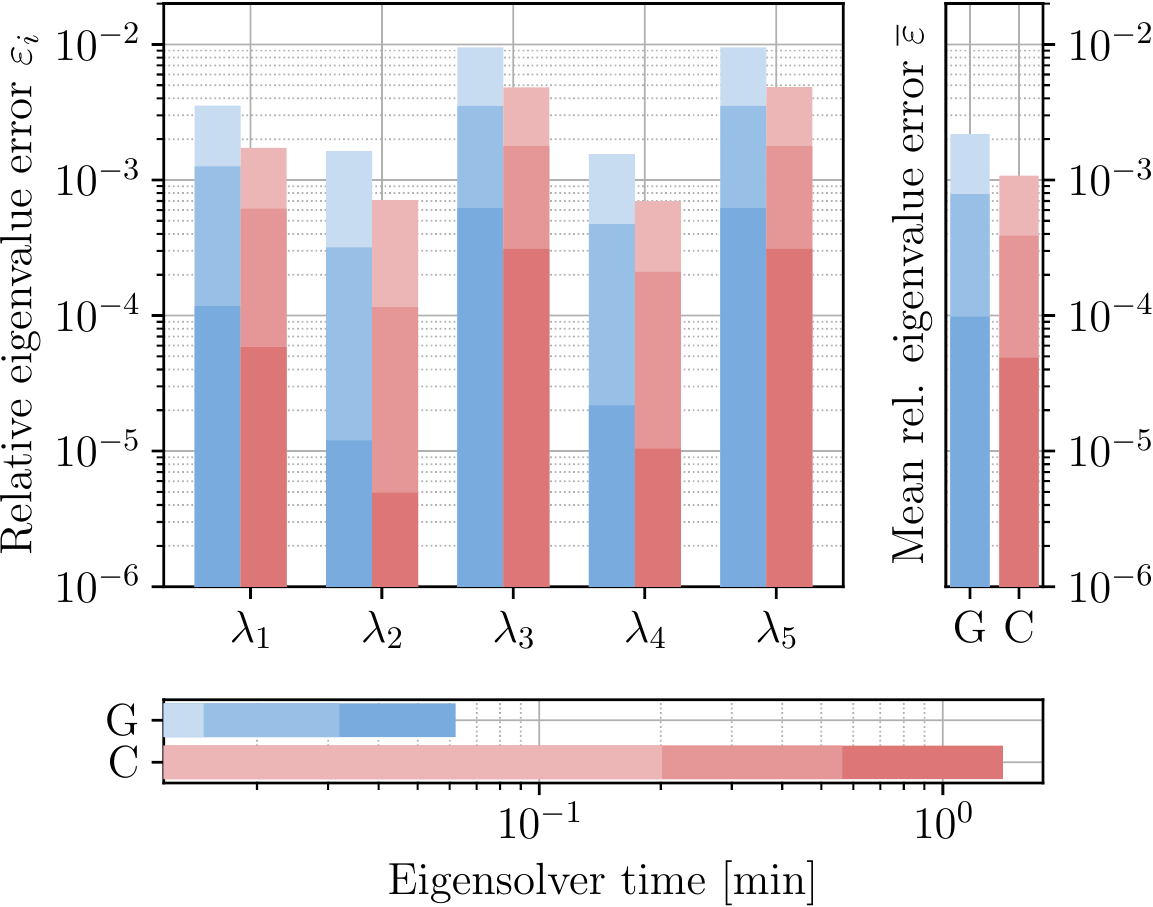} 
\caption{Error of the first five eigenvalues plotted for Cases 1--3 and corresponding timings and accuracy over the first 20 eigenvalues (Example 1, smooth Gaussian kernel).} 
\label{fig:detail-gau} 
\end{figure}

We resort again to the color coding of Table~\ref{tab:colorcoding} to concisely differentiate between the five different resolutions and the two methods.
Figure \ref{fig:scatter-gau} plots the mean relative accuracy of the first twenty eigenvalues versus the eigensolver timings. It is evident that for the smooth Gaussian kernel, the Galerkin method outperforms the collocation method by more than one order of magnitude. Furthermore, in line with the complexity analysis presented in Section \ref{sec:complexity}, we observe that the performance gap increases with increasing polynomial order. Following the scheme of Figure \ref{fig:detail-exp}, we provide a more detailed account of the approximation accuracy of the first five eigenvalues in Figure \ref{fig:detail-gau}.

\section{Conclusions} \label{sec:conclusions}
In this paper, we compared accuracy versus the computational time of two state-of-the-art isogeometric discretization methods for the numerical approximation of the truncated Karhunen--Lo\`eve expansion. The first method is the matrix-free isogeometric Galerkin method proposed by the authors in \cite{mika_matrix-free_2020}. It achieves its computational  efficiency  by combining a non-stan\-dard trial space with a specialized quadrature technique called \emph{interpolation based quadrature}. This method requires a minimum of quadrature points and relies heavily on global sum factorization. The second method is our new matrix-free version of the isogeometric collocation method proposed in \cite{jahanbin_isogeometric_2019}. This method achieves its computational performance by virtue of a low number of point evaluations at collocation points. 

On the one hand, our comparative study showed that for a $C^0$-con\-tinuous exponential kernel, the matrix-free collocation method was about twice as fast at the same level of accuracy as the Galerkin method. On the other hand, our comparative study showed that for a smooth Gaussian kernel, the matrix-free Galerkin method was roughly one order of magnitude faster than the collocation method at the same level of accuracy. Furthermore, the computational advantage of the Galer\-kin method over the collocation method increases with increasing polynomial degree. These results are not surprising, since it was already shown in \cite{mika_matrix-free_2020} that interpolation based quadrature scales virtually independently of the polynomial degree. In our study, we also illustrated via complexity analysis that the matrix-free collocation method scales unfavorably with polynomial order. The suboptimal accuracy of the interpolation based quadrature for rough kernels is also known and was already discussed by the authors in \cite{mika_matrix-free_2020}. Besides the aspect of computational performance, we also showed that both methods are highly memory efficient by virtue of their matrix-free formulation.


As for future work, the advantageous properties inherited by the Galerkin method such as symmetric, positive (semi-)definite system matrices, monotonic convergence of the solution and availability of established mathematical framework for stability and convergence deserve a more detailed theoretical discussion with regard to the interpolation based quadrature method. A generalized accuracy study and more numerical benchmarks with existing methods are desirable as well.


%


\begin{acknowledgements}
D. Schil\-linger gratefully acknowledges funding from the German Research Foundation (DFG) through the Emmy Noether Award SCH 1249/2-1.
\end{acknowledgements}

%
%

\bibliographystyle{unsrt}   

\bibliography{zotero}

\begin{thebibliography}{10}

\bibitem{mika_matrix-free_2020}
M.L. Mika, T.J.R. Hughes, D.~Schillinger, P.~Wriggers, and R.R. Hiemstra.
\newblock A matrix-free isogeometric {Galerkin} method for {Karhunen}-{Loève}
  approximation of random fields using tensor product splines, tensor
  contraction and interpolation based quadrature.
\newblock {\em arXiv:2011.13861 [cs]}, November 2020.

\bibitem{jahanbin_isogeometric_2019}
R.~Jahanbin and S.~Rahman.
\newblock An isogeometric collocation method for efficient random field
  discretization.
\newblock {\em International Journal for Numerical Methods in Engineering},
  117(3):344--369, January 2019.

\bibitem{keese_review_2003}
A.~Keese.
\newblock A {Review} of {Recent} {Developments} in the {Numerical} {Solution}
  of {Stochastic} {Partial} {Differential} {Equations} ({Stochastic} {Finite}
  {Elements}).
\newblock {\em Braunschweig, Institut für Wissenschaftliches Rechnen}, 2003.

\bibitem{stefanou_stochastic_2009}
G.~Stefanou.
\newblock The stochastic finite element method: {Past}, present and future.
\newblock {\em Computer Methods in Applied Mechanics and Engineering},
  198:1031--1051, 2009.

\bibitem{sudret_stochastic_2000}
B.~Sudret and A.~Kuyreghian.
\newblock {\em Stochastic finite element methods and reliability: a
  state-of-the-art report.}
\newblock Berkeley, Department of Civil and Environmental Engineering,
  University of California, 2000.

\bibitem{lu_review_2019}
K.~Lu, Y.~Jin, Y.~Chen, Y.~Yang, L.~Hou, Z.~Zhang, Z.~Li, and C.~Fu.
\newblock Review for order reduction based on proper orthogonal decomposition
  and outlooks of applications in mechanical systems.
\newblock {\em Mechanical Systems and Signal Processing}, 123:264--297, May
  2019.

\bibitem{rathinam_new_2003}
M.~Rathinam and L.R. Petzold.
\newblock A {New} {Look} at {Proper} {Orthogonal} {Decomposition}.
\newblock {\em SIAM Journal on Numerical Analysis}, 41(5):1893--1925, January
  2003.

\bibitem{jolliffe_principal_2016}
I.T. Jolliffe and J.~Cadima.
\newblock Principal component analysis: a review and recent developments.
\newblock {\em Philosophical Transactions of the Royal Society A},
  374(2065):20150202, April 2016.

\bibitem{liang_proper_2002}
Y.C. Liang, H.P. Lee, S.P. Lim, W.Z. Lin, K.H. Lee, and C.G. Wu.
\newblock {Proper} {orthogonal} {decomposition} {and} {its}
  {applications}--{Part} {I}: {Theory}.
\newblock {\em Journal of Sound and Vibration}, 252(3):527--544, May 2002.

\bibitem{eiermann_computational_2007}
M.~Eiermann, O.G. Ernst, and E.~Ullmann.
\newblock Computational aspects of the stochastic finite element method.
\newblock {\em Computing and Visualization in Science}, 10(1):3--15, February
  2007.

\bibitem{saad_numerical_2011}
Y.~Saad.
\newblock {\em Numerical methods for large eigenvalue problems}.
\newblock Number~66 in Classics in applied mathematics. Society for Industrial
  and Applied Mathematics, Philadelphia, rev. ed edition, 2011.

\bibitem{atkinson_numerical_1997}
K.E. Atkinson.
\newblock {\em The {Numerical} {Solution} of {Integral} {Equations} of the
  {Second} {Kind}}.
\newblock Cambridge University Press, 1 edition, June 1997.

\bibitem{ghanem_stochastic_1991}
R.G. Ghanem and P.D. Spanos.
\newblock {\em Stochastic {Finite} {Elements}: {A} {Spectral} {Approach}}.
\newblock Springer New York, New York, NY, 1991.

\bibitem{rahman_galerkin_2018}
S.~Rahman.
\newblock A {Galerkin} isogeometric method for {Karhunen}–{Loève}
  approximation of random fields.
\newblock {\em Computer Methods in Applied Mechanics and Engineering},
  338:533--561, August 2018.

\bibitem{bressan_sum_2019}
A.~Bressan and S.~Takacs.
\newblock Sum factorization techniques in {Isogeometric} {Analysis}.
\newblock {\em Computer Methods in Applied Mechanics and Engineering},
  352:437--460, August 2019.

\bibitem{auricchio_isogeometric_2010}
F.~Auricchio, L.~Beirão Da~Veiga, T.~J.~R. Hughes, A.~Reali, and G.~Sangalli.
\newblock Isogeometric collocation methods.
\newblock {\em Mathematical Models and Methods in Applied Sciences},
  20(11):2075--2107, November 2010.

\bibitem{Schillinger:13.2}
D.~Schillinger, J.A. Evans, A.~Reali, M.A. Scott, and T.J.R. Hughes.
\newblock {I}sogeometric collocation: {C}ost comparison with {G}alerkin methods
  and extension to adaptive hierarchical {NURBS} discretizations.
\newblock {\em Computer Methods in Applied Mechanics and Engineering},
  267:170--232, 2013.

\end{thebibliography}

\end{document}